# AI model GPT-3 (dis)informs us better than humans

Manuscript

**Giovanni Spitale**

**0000-0002-6812-0979**

Institute of Biomedical Ethics and History of Medicine, University of Zurich, Zurich, Switzerland

giovanni.spitale@ibme.uzh.ch

**Nikola Biller-Andorno**

**0000-0001-7661-1324**

Institute of Biomedical Ethics and History of Medicine, University of Zurich, Zurich, Switzerland

biller-andorno@ibme.uzh.ch

**Federico Germani (corresponding author)**

**0000-0002-5604-0437**

Institute of Biomedical Ethics and History of Medicine, University of Zurich, Zurich, Switzerland

federico.germani@ibme.uzh.ch

+41 44 634 40 80

Institute for Biomedical Ethics and History of Medicine (IBME), Winterthurerstrasse 30, 8006 Zürich (CH)

## Abstract

Artificial intelligence is changing the way we create and evaluate information, and this is happening during an infodemic, which has been having dramatic effects on global health. In this paper we evaluate whether recruited individuals can distinguish disinformation from accurate information, structured in the form of tweets, and determine whether a tweet is organic or synthetic, i.e., whether it has been written by a Twitter user or by the AI model GPT-3. Our results show that GPT-3 is a double-edge sword, which, in comparison with humans, can produce accurate information that is easier to understand, but can also produce more compelling disinformation. We also show that humans cannot distinguish tweets generated by GPT-3 from tweets written by human users. Starting from our results, we reflect on the dangers of AI for disinformation, and on how we can improve information campaigns to benefit global health.

## Introduction

GPT-3, the latest iteration of the Generative Pre-trained Transformers by OpenAI, has gained much attention in recent years[1]. It is the most advanced pre-trained language representation system, is a statistical representation of language, and works by producing credible text based on user prompts[2–4]. In fact, an initial test on people's ability to tell whether a ∼ 500 word article was written by humans or GPT-3 showed a mean accuracy of 52%; just slightly better than chance [1].

GPT-3 has remarkable capabilities, but also carries potential implications. It can be a great tool for machine translations, text classification, dialogue/chatbot systems, knowledge summarizing, question answering, creative writing [2,5,6], detecting hate speech [7], and automatic code writing [2,8], but also to produce 'misinformation, spam, phishing, abuse of legal and governmental processes, fraudulent academic essay writing and social engineering pretexting' [1,9–11]. GPT-3 is an amplifier of human intentions which can receive instructions in natural language and its output can be in natural or formal language. This tool is neutral from an ethical point of view – and it is subject to the dual use problem[12].

The advancements in AI text generators and the release of GPT-3 coincide with the ongoing infodemic[13] – an epidemic-like circulation of fake news and disinformation, which, alongside the COVID-19 pandemic, has been greatly detrimental for global health. Given GPT-3's potential misuse and impact on global health, it is crucial to evaluate how its text affects people's understanding of information.

In this paper we aim to determine whether GPT-3 can be used to produce accurate information and disinformation in the form of tweets, and compare its credibility to that of human-generated information. At the same time we aim to determine whether GPT-3 can be used to develop assistive tools to help identifying disinformation. We acknowledge that the definitions of disinformation and misinformation are diverse; in this paper we refer to an inclusive definition, which considers as disinformation both false information (also partially false information) and/or misleading content[14].

To achieve our goals, we asked GPT-3 to write information or disinformation tweets on topics prone to disinformation and public misconception, such as vaccines, 5G, COVID-19, or the theory of evolution. We collected real tweets on these topics and created a survey in which participants were asked to classify synthetic tweets (written by GPT-3) and organic tweets (written by humans) as true or false and identify if they were written by a real user or AI.

## Results

### Study design and demographics

To test GPT-3's ability to generate accurate or fake tweets, we created prompts instructing GPT-3 to generate tweets with either accurate information or disinformation on the following topics: climate change, vaccines safety, theory of evolution, COVID-19, masks safety, vaccines and autism, homeopathy treatments for cancer, flat Earth, 5G technology and COVID-19, antibiotics and viral infections, and COVID-19 and influenza. We performed a Twitter search to identify accurate tweets and disinformation tweets generated by Twitter users. We call "synthetic" those tweets that are generated by GPT-3, and we call "organic" those real tweets retrieved from Twitter. Human respondents were recruited online to participate in a quiz, in which they were asked to recognize whether a set of tweets were organic or synthetic, and true or false (i.e., whether they contained accurate information or disinformation). GPT-3 was also questioned about whether tweets forming the same dataset were true or false (Figure 1A). We recruited 869 respondents. 157 responses were excluded because they were incomplete. 15 additional responses were removed because the respondents were too fast to meaningfully complete the survey, for a total of 697 responses included in our analysis (Figure 1B). Most of the respondents were from the UK, Australia, Canada, USA, and Ireland (Figure S1A), with more females than males (Figure S1B), a balanced age, with a high representation of people between 42 and 76 years old (Figure S1C), and a balanced education level profile, with most of the respondents holding a Bachelor's degree (Figure S1D,E).

## GPT-3 AI model informs and disinforms us better

We measured how accurately participants recognized whether a tweet was containing disinformation or accurate information (Disinformation Recognition Score), for four types of tweet: "Organic true", which are tweets published by Twitter users (organic) and containing accurate information (true); "Synthetic true", which are tweets generated by GPT-3 (synthetic) and containing accurate information (true); "Organic false", tweets generated by Twitter users (organic) and containing disinformation (false); and finally "synthetic false", tweets generated by GPT-3 (synthetic) and containing disinformation (false). Participants recognized "organic false" tweets with higher accuracy, compared with synthetic false tweets (Scores 0.92 versus 0.89, respectively; p=0.0032) (Figure 1C). Similarly, they recognized "synthetic true" tweets correctly more than "organic true" tweets (Scores 0.84 versus 0.72, respectively; p<0.0001). This indicates that human respondents can recognize the accuracy of tweets containing accurate information more often when such tweets are generated by GPT-3, when compared with organic tweets retrieved from Twitter. Similarly, GPT-3 generated disinformation tweets are more successful in misleading when compared with organic ones, albeit with a small effect. When evaluating the same dataset, segmenting the analysis for true versus false tweets – regardless of whether they are organic or synthetic – and for organic versus synthetic tweets – regardless of their truthfulness – tweets containing accurate information received lower scores when compared with tweets containing disinformation (Scores 0.78 versus 0.91, respectively; p<0.0001). Similarly, synthetic tweets were categorized more often correctly for the accuracy of the information they contained (Scores 0.87 versus 0.82, respectively; p<0.0001) (Figure 1C'). Participants required on average 29.14 seconds to determine whether an "organic true" tweet was accurate or contained disinformation. This was significantly more when compared with "organic false" tweets, which required 23.28 seconds for evaluation, with "synthetic true" tweets requiring 21.02 seconds, and "synthetic false" tweets requiring 19.87 seconds (Figure 1D). True tweets required a longer time for evaluation when compared with false tweets (25.07 vs 21.97 seconds, p<0.0001), as well as organic tweets when compared with synthetic tweets (26.21 vs 20.44 seconds, p<0.0001). (Figure 1D'). The time required for evaluation was not dependent on the length of tweets (Figure 1E). Further, we calculated Disinformation Recognition Scores for each category (e.g., "climate change", "vaccines and autism"), for each type of tweet (i.e., "organic true", "synthetic true", "organic false", "synthetic false") (Figure S2), and plotted the average Disinformation Scores for each type of tweet (Figure S3), obtaining comparable results with the analysis run on the Disinformation Recognition Scores of each respondent. This confirms that humans find it harder to assess accurate information than disinformation and GPT-3 is more efficient in informing and misleading humans, as respondents can evaluate its tweets faster, correctly or incorrectly. A list of the disinformation tweets recognized most often as accurate tweets can be seen in Figure S4, and a list of tweets containing accurate information, recognized most often as disinformation tweets, can be seen in Figure S5.

## Humans evaluate the accuracy of information better than GPT-3

The respondents of our survey evaluated the accuracy or inaccuracy of the information contained in 220 tweets. Using the same dataset, we asked GPT-3 to evaluate whether the tweets were accurate or whether they contained disinformation. For disinformation tweets, humans and GPT-3 performed similarly, although respondents performed slightly better (Success rates: 0.92 vs 0.89, respectively). For accurate tweets, GPT-3, likewise human respondents, had more difficulties evaluating the accuracy of the information. In comparison, human respondents performed better than GPT-3 (Success rates: 0.72 vs 0.64, respectively) (Figure 2A). A detailed analysis of the results for each category of tweets can be found in Figure S6. These results suggest that human respondents can evaluate information better than GPT-3. Considering that these respondents are not necessarily trained individuals in recognizing disinformation, trained humans, at the time of writing, may perform much better than machines at performing this task.

## GPT-3 can "disobey" requests to produce disinformation

GPT-3 lacks mental representations and intentionality[15,16], therefore using quotes for "obedience" and "disobedience" is necessary. As mentioned, we instructed GPT-3 to produce a set of true and false tweets (i.e., accurate or disinformation tweets) (Figure 1A). We instructed GPT-3 to produce 10 accurate and 10 disinformation tweets for each category. Of these, we included in our survey only the tweets for which GPT-3 "obeyed" our request to produce accurate or disinformation tweets, respectively. We calculated the rate of obedience, i.e., the percentage of requests that were satisfied by GPT-3 divided for the overall number of requests we made to GPT-3. For accurate information, GPT-3 obeyed our requests 99 times out of 101, whereas for disinformation the rate of obedience was much lower (80/102) (Figure 2B), indicating that GPT-3 can "refuse" to produce disinformation, and in rarer instances, it may produce disinformation when asked to produce accurate information. For a detailed analysis of the obedience rate per category of tweet, we refer to Figure S7.

## Information generated by GPT-3 is undistinguishable from human-generated information

We calculated the AI Recognition Score (or OS score), i.e., the ability of respondents to recognize whether tweets are organic (i.e., produced by Twitter users) or synthetic (i.e., produced by GPT-3). As for the evaluation of the ability to recognize disinformation, we calculated the AI Recognition Score for each type of tweet (i.e., "organic true", "synthetic true", "organic false", "synthetic false"). Respondents scored around 0.5 on average, suggesting random choice between synthetic and organic tweets and inability to distinguish AI-generated tweets from real ones (Figure 3). That said, respondents obtained a significantly higher score for "organic true" tweets and, to a lesser extent, for "organic false" tweets (0.67 and 0.60, respectively), whereas for "synthetic true" and "synthetic false" tweets, scores were below 0.5 (0.34 and 0.40, respectively) (Figure 3A). Looking at true versus false tweets, and organic versus synthetic tweets, we note something interesting; people's ability to determine organic versus synthetic tweets is not influenced by tweet accuracy, with answers being random on average. (0.50 vs 0.50, respectively, p=0.9576). However, when comparing organic versus synthetic tweets, organic tweets were recognized more often as organic when compared with synthetic tweets recognized as synthetic (0.63 vs 0.37, p<0.0001) (Figure 3A'). Therefore, both organic and synthetic tweets tend to be classified as "human", indicating that GPT-3 can effectively mimic human-generated information. Further, we calculated AI Recognition Scores for each category (e.g., "climate change", "vaccines and autism"), for each type of tweet (i.e., "organic true", "synthetic true", "organic false", "synthetic false") (Figure S8), and plotted the average AI Recognition Scores for each type of tweet (Figure S9), obtaining comparable results with the analysis run on the AI Recognition Scores of each respondent. A list of the organic tweets recognized most often as synthetic can be seen in Figure S10, and a list of synthetic tweets recognized most often as organic can be seen in Figure S11.

## Building versus crashing confidence

At the beginning of the survey, we asked respondents to define how confident they were in their ability to recognize disinformation, and in their ability to identify AI versus human-generated text, using a 1 to 5 Likert scale (Figure 4A,B). The Disinformation Recognition Confidence before the test was higher than AI Recognition Confidence before the test (3.05 versus 2.69, respectively). After taking the survey (but before seeing their results), we asked again respondents to define how confident they were in their ability to recognize disinformation and AI versus human-generated text. Respondents were more confident in their ability to recognize disinformation (Pre versus Post, 3.05 versus 3.49, respectively, p<0.0001) (Figure 4A), whereas they were much less confident in their ability to recognize synthetic versus organic tweets (Pre versus Post, 2.79 versus 1.70, respectively, p<0.0001) (Figure 4B). The increase in confidence to recognize disinformation could be explained by the inoculation theory of misinformation[17], whereby critical exposure to disinformation could improve disinformation recognition capacity and resilience. Instead, the stark decrease in confidence to recognize synthetic tweets could depend on what we could call "resignation theory" – i.e., people may abandon their efforts to critically assess information when faced with a vast amount of confusing information. This could result in apathy and reliance on emotions for information consumption.

## Figures

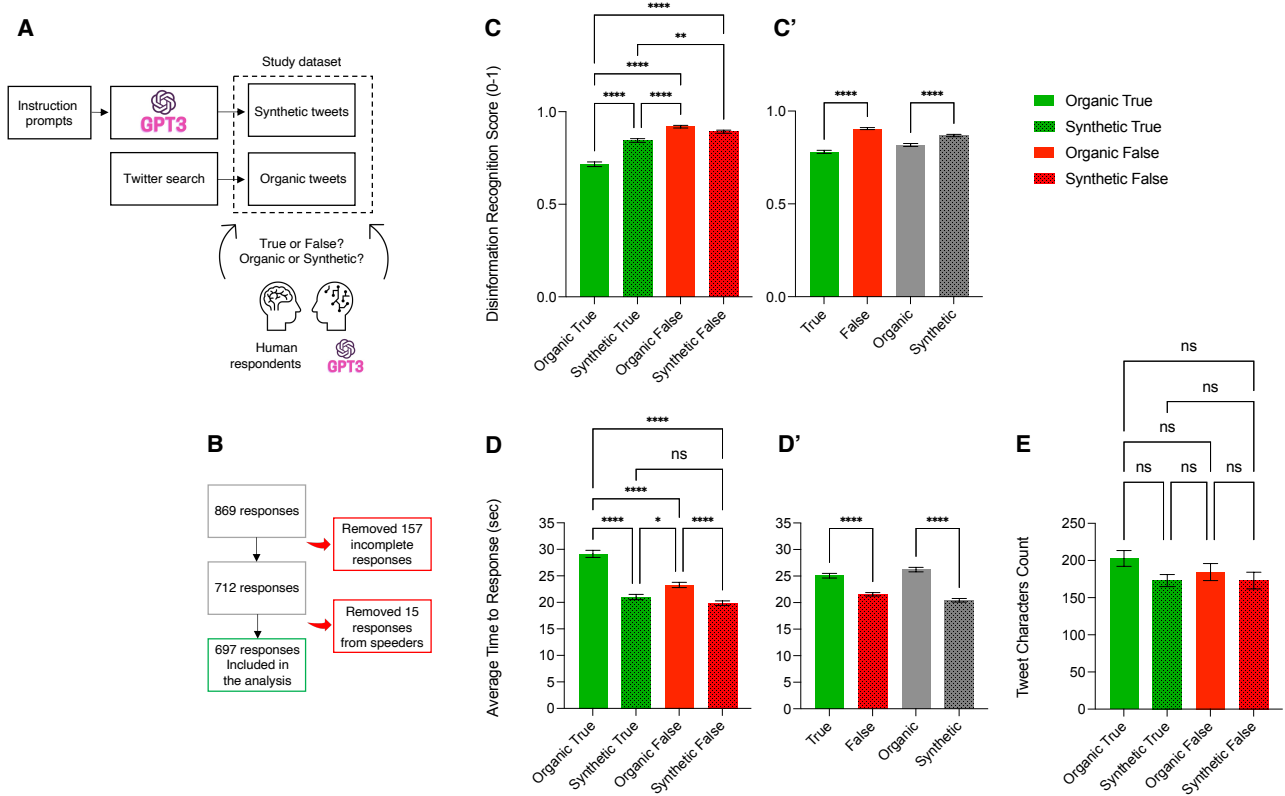

**Figure 1. GPT-3 AI model informs and disinform us better.** Study set-up: GPT-3 was provided instruction prompts to produce synthetic tweets containing accurate versus inaccurate information (information vs disinformation). Through a Twitter search, organic tweets from Twitter users were retrieved and classified as accurate information or disinformation. The sum of synthetic and organic tweets, either containing information or disinformation, constitutes the study dataset. Human respondents were asked to recognize whether such tweets were true or false (i.e., accurate information or disinformation) and whether the tweets were organic (i.e., generated by Twitter users) or synthetic (i.e., generated by GPT-3). GPT-3 was asked to recognize whether the tweets were true or false (i.e., accurate information or disinformation) (**A**). Data collection: we gathered 869 responses to our survey. 157 responses were incomplete and were removed. Of 712 remaining responses, 15 were removed as they were completed too fast to be reliable. Our analysis was conducted on 697 complete and reliable responses (**B**). GPT-3 produces accurate and disinformation tweets that are recognized by human respondents as accurate more often than accurate and disinformation tweets, respectively, produced by humans. "Organic true" tweets (green column bars) are accurate tweets generated by Twitter users; "Synthetic true" (dotted green column bars) tweets are accurate tweets generated by GPT-3; "Organic false" (red column bars) tweets are disinformation tweets generated by Twitter users; "Synthetic false" (dotted red column bars) tweets are disinformation tweets generated by GPT-3 (**C**). Disinformation tweets (red column bars) are recognized more often correctly when compared with accurate tweets (green column bars). Synthetic tweets (dotted grey column bars) are recognized more often correctly when compared with organic tweets (grey column bars). Disinformation recognition score (or TF score) (0-1) is the average score for all 697 respondents (1= 100% correct answers; 0 = 0% correct answers); Ordinary one-way ANOVA multiple-comparisons Tukey's test, n=697, ** p<0.01;

****p<0.0001. Error bars = SEM. (**C'**). Average time to response in seconds for "organic true", "synthetic true", "organic false" and "synthetic false" tweets. Organic true tweets required the most time to receive an evaluation from respondents, whereas synthetic true and synthetic false tweets required the least time to receive an evaluation (D). True tweets (i.e., accurate) required longer to be evaluated by respondents than false tweets (i.e., containing disinformation), and organic tweets required more time to be evaluated when compared with synthetic tweets. Ordinary one-way ANOVA multiple-comparisons Tukey's test, n=697, ns = non-significant, p>0.05; *p<0.05; ****p<0.0001. Error bars = SEM. (**D'**). The different time required to evaluate each type of tweet does not depend on the length of tweets, since the tweet characters count is no different between types. Ordinary one-way ANOVA multiple-comparisons Tukey's test, n(total)=220, n(type)= 55; ns = non-significant, p>0.05 (**E**).

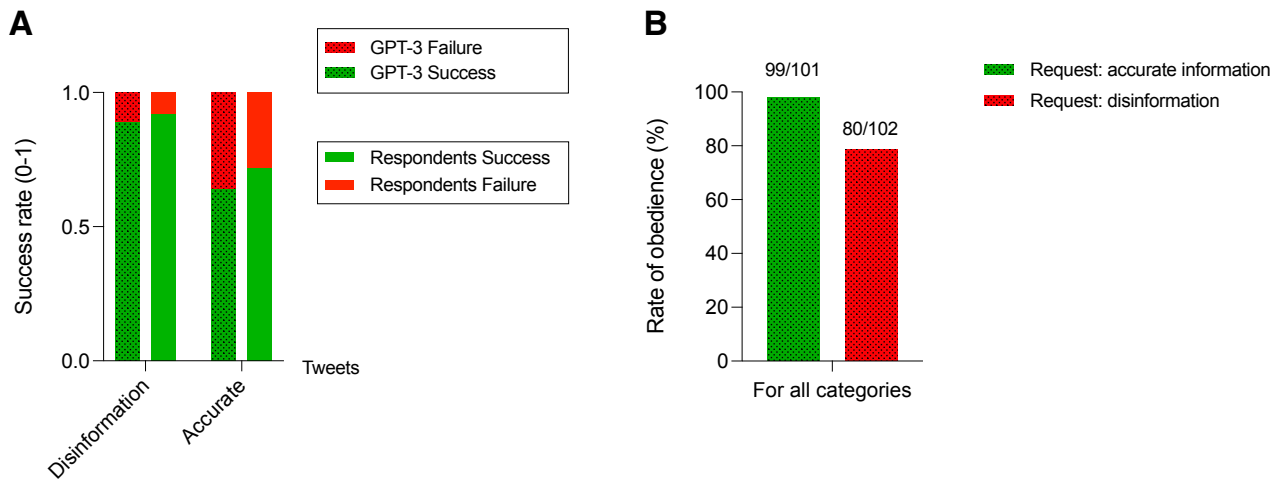

**Figure 2. Humans evaluate information and disinformation better than GPT-3, and GPT-3 can "disobey" requests to generate disinformation.** Green column bars represent successful responses given by human respondents, whereas green dotted bars represent successful responses given by GPT-3. Red bars represent incorrect responses from human respondents, whereas red dotted bars represent incorrect responses from GPT-3. The success rate concerning the evaluation of disinformation is 89% and 92% for GPT-3 and human respondents, respectively. The success rate concerning the evaluation of accurate information is 64% and 72% for GPT-3 and human respondents, respectively. The evaluation was conducted on organic tweets retrieved from Twitter which were included in our survey (**A**). Rate of "obedience" for GPT–3 – i.e., how often GPT-3 respected our request to generate information or disinformation tweets. For accurate information tweets, GPT-3 "obeyed" our request 99 times out of 101 requests, whereas for disinformation tweets, it "obeyed" our request 80 times out of 102 requests (**B**).

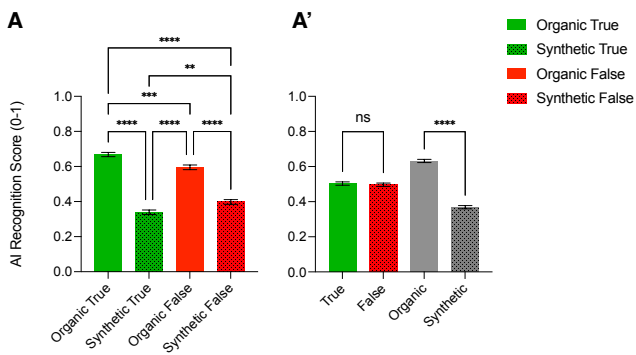

**Figure 3. Human respondents cannot distinguish organic versus synthetic tweets, but recognize their origin better when they are generated by humans.** AI recognition score for "organic true" (green bars), "synthetic true" (green dotted bars), "organic false" (red bars) and "synthetic false" (red dotted bars) tweets. AI recognition score (0-1) indicates the probability that human respondents can identify whether a tweet is produced organically (i.e., by a Twitter user), or synthetically (i.e., by GPT-3). Human respondents recognize whether organic true tweets are organic or synthetic tweets more effectively than all other type of tweets, whereas synthetic true tweets are recognized correctly the least (**A**). Human respondents cannot predict whether true or false tweets (i.e., accurate tweets or disinformation tweets, green versus red bars) are produced by humans or by GPT-3, and the truthfulness of the information does not have an impact on the AI recognition score. Regarding organic versus synthetic tweets (grey versus grey dotted bars), human respondents recognize whether tweets are generated by humans or GPT-3 better when they are organic (i.e., generated by Twitter users), when compared with synthetic tweets (i.e., generated by GPT-3). AI recognition score (0-1) is the average score for all 697 respondents (1= 100% correct answers; 0 = 0% correct answers); Ordinary one-way ANOVA multiple-comparisons Tukey's test, n=697; ns = non-significant, p>0.05; **p<0.01; ***p<0.001; ***p<0.0001. Bars represent SEM.

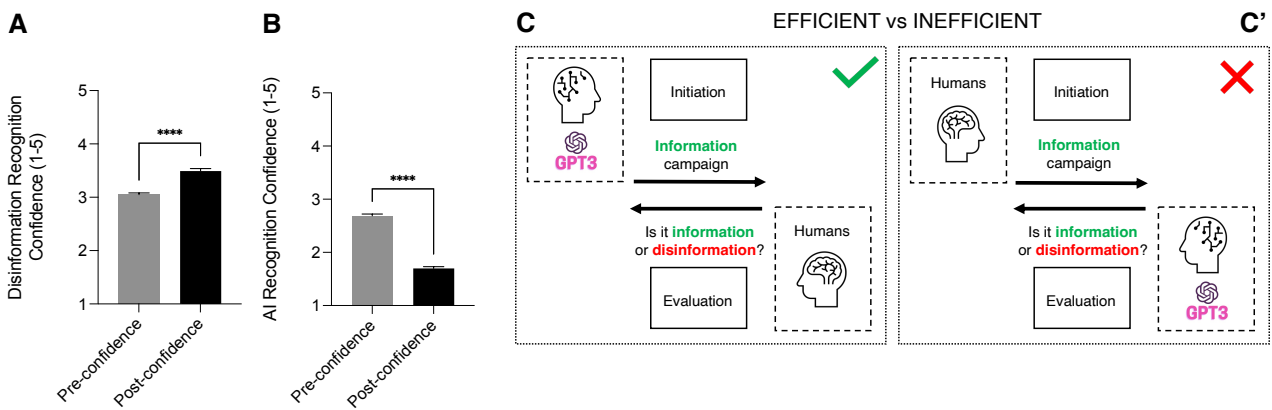

**Figure 4. The Confidence in recognizing disinformation increases post-survey, whereas the confidence in recognizing AI-generated information decreases; and proposed model to launch information campaigns and evaluate information.** Respondents were asked to provide a score of how confident they were in their ability to recognize disinformation tweets before taking the survey (grey bar), and after taking the survey (black bar). Participants' confidence in disinformation recognition increased significantly from 3.05 to 3.49 out of 5. n=697; Welch's t-test, ****p<0.0001. Bars represent SEM (**A**). Respondents were asked to provide a score of how confident they were in their ability to recognize whether tweets were generated by humans (grey bar) or by AI (black bar). Participant's confidence in AI recognition dropped significantly from 2.69 to 1.7 out of 5. n=697; Welch's t-test, ****p<0.0001. Bars represent SEM (**B**). Model for an efficient and inefficient communication strategy and launch of information campaign. Based on our data, and with the AI model adopted for our analysis, an efficient system relies on accurate information generated by GPT-3 (initiation phase), whereas it relies on trained humans to evaluate whether a piece of information is accurate or whether it contains disinformation (evaluation phase) (**C**). An inefficient system relies on humans to generate information and initiate an information campaign, and it relies on AI to evaluate whether a piece of information is accurate or whether it contains disinformation (**C'**).

# Discussion

## How to communicate and evaluate information

Our findings show that tweets produced by GPT-3 can both inform and disinform better than organic tweets. Synthetic tweets containing reliable information are recognized as true better and faster than true organic tweets, while false synthetic tweets are recognized as false worse than false organic tweets. GPT-3 does not perform better than humans in recognizing both information and disinformation. The results suggest GPT-3 may be better at informing because it can generate easier-to-read and understand text than humans. Based on these results, we propose a model for efficient communication and evaluation of information. Instead of humans creating information and AIs evaluating it [18] (Figure 4C, C'), well-structured prompts can be given to GPT-3 to create effective information campaigns (initiation phase). The accuracy of information is then evaluated by trained humans (Figure 4C). Instead, information campaigns written and prepared by humans would turn out to be less effective, and AIs would perform an inefficient evaluation of how truthful and reliable information is (Figure 4C'). This model is relevant in a public health crisis and infodemic as it allows for fast and clear communication to the public.

## "Disobedience", training datasets, and error propagation

Our results show that, when asked, GPT-3 is less likely to produce misinformation on some specific topics, e.g.: vaccines and autism (Figure S7). Being GPT-3 a statistical representation of language – for how language is used in the datasets it was trained on – we assume GPT-3's "disobedience" depends on the composition of GPT-3's training datasets. If the training dataset contains volumes of information contradicting what the prompt asks for, the system will likely output that type of information. We can therefore assume that the volume of information in the training dataset debunking causal links between vaccines and autism is higher than the volume of information debunking conspiracy theories on other topics taken into consideration by our study. GPT-3 is trained on data obtained from Common Crawl, WebText2, Books1, Books2 and Wikipedia[19], which could also include misinformation and disinformation. The risk of disinformation can be reduced by training it on regulated datasets that enhance accuracy and transparency, and report their sources. This would allow for independent fact checking.

## 'As human as humans': synthetic text identification and impersonation

In line with previous research[20], we found that both human respondents and GPT-3 struggle to differentiate between organic and synthetic tweets (data on GPT-3's assessment are available in our study's repository)[21]. It might be possible to develop specific training courses to improve humans' recognition of synthetic text, based on linguistic markers, grammatical structure, and syntax, but the availability of ChatGPT, a more interactive and conversational interface for GPT-3, has led to efforts to bypass OpenAI's content policy. An effective strategy is impersonation: when GPT-3 refuses to generate output that could violate the content policy, users just ask it to impersonate a character[22]. With this approach, swathes of disinformation could be produced by first asking GPT-3 to generate fake profiles of people to impersonate, and in a second iteration, to generate tweets that these profiles could write. Besides circumventing content policy blocks, this would add even more 'human-like' feel to the tweets, and make it even harder to identify them as synthetic. Based on these premises, synthetic text identification might soon be a hopeless battle to fight, for both people and AIs.

## Resignation theory

This decrease in self-confidence to distinguish synthetic text from organic text after exposure to both synthetic and organic texts may be due to the realization that there is no clear marker that allows users to identify whether a text has been generated by a machine or a human. This is likely because of GPT-3's ability to mimic human writing styles and language patterns. Additionally, respondents may have initially underestimated GPT-3's abilities to write human-like text: this may be due to the fact that such technology is new and revolutionary, and people are not yet accustomed to how powerful it can be. Another possible interpretation is that the survey may have made participants more aware of GPT-3's potential to generate disinformation with a human-like feel, making them more sceptical of both synthetic and organic information.

## Beyond Twitter

We decided to focus our study on tweets, since Twitter users consume mostly news and political information [23,24]. Also, Twitter offers a very simple application programming interface (API) to develop bots, i.e.: programs able to post content and interact with posts or users without human supervision[25]. Recent research shows that only about 5% of Twitter users are bots – but that these bots cumulatively account for 20% - 29% of the content posted on Twitter[26]. Because of these characteristics, Twitter is the ideal target – and potentially a very vulnerable one – for AI-generated swathes of disinformation. It is important to note that while we generated tweet-like social media posts, our "tweets" have features shared in other type of social media posts, such as Instagram or Facebook posts. Therefore, our results could be extended to other social media platforms and other forms of communication that can be used by bots via APIs, and that could be exploited to programmatically disseminate AI-generated disinformation.

## The genie is out of the bottle

Our findings predict that advanced AI text generators can impact information dissemination positively and negatively. To mitigate negative effects, taking action to regulate which training datasets are used to develop these technologies is crucial, thus ensuring transparency, truthfulness of the output information, and limiting misuse of the technology to generate deceiving information. Until we do not have efficient strategies for identifying disinformation (whether based on human skills or on future AI improvements), restricting the use of these technologies may be necessary, e.g.: licensing them only to trusted users (e.g., research institutions), or limiting the potential of AIs to certain type of applications. Finally, it is crucial that we continue to critically evaluate the implications of these technologies and take action to mitigate any negative effects on society.

# Methods

We registered the protocol of this study before starting the data collection. The pre-registration is available on OSF: https://doi.org/10.17605/OSF.IO/HV6ZY .

## Definition of the topics

As the focus of this study, we initially identified 14 topics on which disinformation exists. This preliminary list included:

- Climate change;
- Vaccines safety;
- Theory of evolution;
- COVID-19;
- Masks safety;
- Vaccines and autism;
- Homeopathic treatments for cancer;
- Flat Earth;
- 5G technology and COVID-19;
- Bill Gates and population control;
- Antibiotics and viral infections;
- COVID-19 = influenza;
- Inferior human races;
- Moral AI.

## Generation of synthetic tweets

Based on the list defined above, we generated synthetic tweets passing input to GPT-3 via API (Application Programming Interface). The code asks to generate 10 true tweets and 10 false tweets for each of the topics detailed above (e.g.: prompt: 'Write a tweet to explain why climate change is real', category: 'Climate change'). The tweet generation code consists of one function to pass input prompts to GPT-3, and of two different loops to iterate over categorized prompts. The first function defines the parameters to pass to GPT-3 (temperature, max_token, top_p, best_of, frequency_penalty, presence_penalty), empirically defined in an iterative process as the most apt to produce text that resembles social media content. GPT-3's API returns also the reason for termination (e.g.: reaching the length specified in max_tokens). For these cases, the text sometimes contains unfinished sentences: these have been removed. The loops to generate true and false tweets read input organized in .csv files (prompt and category) and generate the given number of texts per each prompt (in this example, 10). The output is then exported as a .xlsx file containing three columns: the text, the reason for termination, and the category. All the code, available in this study's pre-registration repository, is organized in commented Jupyter lab notebooks for scrutiny and replication[1]. The prompts and the output are available in the same repository.

## Definitions

Throughout the manuscript we adopt – and sometimes explain for added clarity – the terminology "true" and "false" tweets. True tweets are those tweets containing accurate information, and false tweets are those containing inaccurate information, i.e., disinformation.

As for the definition of accurate information and disinformation, we base ourselves on the current scientific knowledge and understanding of the topics and information under scrutiny. To avoid dubious and debatable cases, which may be subject to personal opinions and interpretations, we only analyzed and added to our questionnaire those tweets containing information that is clearly categorizable as true or false. Of note, if a tweet contained partially incorrect information – meaning it contained more than 1 pieces of information, and at least one was incorrect, it was labelled as false. As discussed in the introductory section of the manuscript, we acknowledge that the definition of disinformation and misinformation is diverse, but we refer to an inclusive definition, which considers false information (also partially false information) and/or misleading content[2].

## Retrieval of organic tweets

Using Twitter's advanced search, we collected a random sample of recent organic tweets on the topics listed above, including both true and false tweets. Our initial aim was to collect 50 tweets per category; however, this proved impossible for some categories, for various reasons – for example, for some categories tweets were ambiguous and difficult to categorize as true or false. For other categories, we were not able to retrieve enough tweets. Some categories were therefore dropped and excluded from the following phases of the study.

| Category | Organic tweets retrieved |
|---|---|
| Climate change | 50 |
| Vaccines safety | 50 |
| Theory of evolution | 49 (1 duplicated tweet was excluded) |
| COVID-19 | 50 |
| Masks safety | 50 |
| Vaccines and autism | 50 |
| Homeopathic treatments for cancer | 30 |
| Flat Earth | 50 |
| 5G technology and COVID-19 | 50 |
| Bill Gates and population control | 0 (dropped) |
| Antibiotics and viral infections | 50 |
| COVID-19 = influenza | 50 |
| Inferior human races | 10 (dropped) |
| Moral AI | 0 (dropped) |

*Table 1. Organic tweets retrieved, by category*

The tweets are available in the study's repository[1].

## Expert assessment of synthetic and organic tweets

We evaluated synthetic and organic tweets to assess whether they contained disinformation. The expert assessment was performed independently by FG and GS, and a following joint analysis was conducted by FG and GS to verify the correctness of their initial assessments.

## Selection of the tweets to include in the survey and generation of tweet images

Based on the assessments defined above, we selected the following tweets for each category:

- 5 synthetic false;
- 5 synthetic true;
- 5 organic false;
- 5 organic true.

We selected only tweets for which FG and GS agreed in their evaluation, following the expert assessment phases. This resulted in a dataframe of 220 tweets (available in our repository[1]) used to generate the images of the tweets. The code generates a random pseudonym and a random username for each tweet (e.g.: 'John S.', @john_s), and creates an image which resembles the screenshot of a tweet. The code, the dataframe containing the tweets, and the output images are available in the study's repository[1].

## AI assessment of tweets

The AI assessment was performed by GPT-3 (true/false evaluation and organic/synthetic evaluation). The first evaluation function defines the parameters to pass to GPT-3 to produce a 'true/false evaluation' (i.e.: whether the tweet is true or false). The second evaluation function defines the parameters to pass to GPT-3 to produce an 'organic/synthetic evaluation' (i.e.: whether the tweet was written by a person or by an AI). The loops for evaluation read the content of the files containing the tweets and evaluate them. The output is scored (i.e.: whether GPT-3's assessment matches the expert assessment for true/false and whether it matches the origin of the tweet for the organic/synthetic classification) and then exported as a .xlsx file. The code and the files containing the assessments are available in the study's repository[1].

## Programming of the survey

We programmed a Qualtrics survey to collect demographics, display the tweets to the respondents, and collect their assessments (true versus false, organic versus synthetic). For each tweet, respondents assess:

- Whether it is accurate or contains disinformation (single choice, accurate/misinformation);
- Whether it was written by a real person or generated by a computer (single choice, real person/computer).

Additionally, respondents provide:

- Some demographic information (nationality, age, sex, education level, education field)
- Self-perceived (pre and post survey) ability to recognize, respectively, disinformation and synthetic text (Likert scale, 1 - very difficult - 5 - very easy)

The images of the tweets are organized in nested randomizers within the survey structure:

- the first level randomizer randomizes the category order (climate change, ...). All the categories are displayed to every respondent.
- second level randomizers (for each category) randomize the single tweet displayed for each category to the respondent. Each category comprises a total of 20 tweets: 5 synthetic false, 5 synthetic true, 5 organic false, 5 organic true. The second level randomizers evenly present one tweet from the pool of 20 tweets.

The survey adopts a gamified approach to keep respondents engaged: at the beginning of the survey, respondents are told that, upon competition, they will obtain their score for both scales (disinformation recognition, and synthetic text recognition). This ensured a low dropout rate. In-survey scoring is achieved using the 'scoring' function in Qualtrics. The survey file and structure are available in the study's repository[1].

## Pilot testing

We pilot tested the survey in two phases. During the first phase we circulated the link to a convenience sample with the aim to test the usability and the layout. This led to minor modifications in the interface and in the wording. During the second phase we distributed the link via a Facebook ads campaign, structured as follows:

- Daily budget: 15€
- Start: 04.10.2022
- End: 13.10.2022
- Age: 16 - 65+
- Languages: English
- Title: True or False? Organic or synthetic?

- Description: Are you able to distinguish text written by an artificial intelligence from text written by a human being? And accurate information from misinformation? Find out with this test, and contribute to research on information ethics.
- Image: generated by DALL·E 2 (available in this study's repository[1])

The campaign had a total cost of 122.92€. It generated a total of 593 clicks on the link (cost per click: 0.21€) and a total of 276 responses (cost per response: 0.46€). The campaign was launched and completed in October 2022.

## Sample size and power analysis
Based on pilot data, we conducted a power analysis to determine the sample size for the full study.

### Primary endpoint hypothesis
Disinformation produced by a machine is more credible than disinformation produced by a human (synthetic versus organic disinformation).

### Secondary endpoints hypotheses
1. Accurate information produced by a machine is more credible than accurate information produced by a human (synthetic versus organic accurate information).
2. Users recognize and distinguish information produced by humans and by machines (regardless of the truthfulness of the information).
3. The confidence of respondents in recognizing disinformation increases after the completion of the questionnaire.
4. The confidence of respondents in recognizing synthetic versus organic information increases after the completion of the questionnaire.

### Power analysis
Based on the data resulting from the pilot study, available in the study's repository[1], we performed a power analysis to estimate the sample size necessary to draw sufficiently meaningful conclusions for Primary and Secondary Endpoints (PE and SEs). Endpoints are continuous, and the study runs on two independent samples.

#### *Primary endpoint: Results*
Average group 1 Score (Synthetic tweets, disinformation)* = 0.86

* From 0 to 1, the score indicates how good the performance was in recognizing synthetic tweets containing disinformation.

Stdev group 1 = 0.25

Average group 2 Score (Organic tweets, disinformation)* = 0.89

* From 0 to 1, the score indicates how good the performance was in recognizing organic tweets containing disinformation.

Enrollment ratio = 1.01194

Alpha = 0.05 Power = 80%

**Sample Size Total = 2181** (Group 1: 1084; Group 2: 1097)

#### *Secondary Endpoint 1*
Average group 1 Score (Synthetic tweets, accurate information)* = 0.78

* From 0 to 1, the score indicates how good the performance was in recognizing synthetic tweets containing accurate information.

Stdev group 1 = 0.35

Average group 2 Score (Organic tweets, accurate information)* = 0.64

* From 0 to 1, the score indicates how good the performance was in recognizing organic tweets containing accurate information.

Enrollment ratio = 0.991045

Alpha = 0.05 Power = 80%

**Sample Size Total = 197** (Group 1: 99; Group 2: 98)

*Secondary Endpoint 2*
Average group 1 Score (Synthetic tweets [accurate information + disinformation])* = 0.315

* From 0 to 1, the score indicates how good the performance was in recognizing synthetic tweets, regardless of whether they contained accurate information or disinformation.

Stdev group 1 = 0.44

Average group 2 Score (Organic tweets, [accurate information + disinformation])* = 0.59

* From 0 to 1, the score indicates how good the performance was in recognizing organic tweets, regardless of whether they contained accurate information or disinformation.

Enrollment ratio = 1.001493

Alpha = 0.05 Power = 80%

**Sample Size Total = 80** (Group 1: 40; Group 2: 40)

*Secondary Endpoint 3*
Average group 1 Score (Pre-confidence level in ability to recognize disinformation)* = 2.932271

* From 1 to 5

Stdev group 1 = 0.829093

Average group 2 (Post-confidence level in ability to recognize disinformation)* = 3.319149

* From 1 to 5

Enrollment ratio = 1.0680

Alpha = 0.05 Power = 80%

**Sample Size Total = 145** (Group 1: 70; Group 2: 75)

*Secondary Endpoint 4*
Average group 1 (Pre-confidence level in ability to recognize synthetic versus organic contents)* = 2.703557

* From 1 to 5

Stdev group 1 = 0.897012

Average group 2 (Post-confidence level in ability to recognize synthetic versus organic contents)* = 1.75

\* From 1 to 5

Enrollment ratio = 1.0720

Alpha = 0.05 Power = 80%

**Sample Size Total = 27** (Group 1: 13; Group 2: 14)

## Sample size evaluation

Taking the larger sample size resulting from our power analyses (n=2181 assessments for PE), and considering that we obtained 1348 assessments (organic, disinformation + synthetic, disinformation), and considering that the pilot study has generated full responses from 277 respondents, the ratio between target power (number of assessments) and sample size of the pilot study (number of assessments) is 1.617953. Therefore, the number of users required for the study is 277*1.617953 = 448.1728. We established that the minimum number of respondents to achieve a properly powered analysis in the full study is n=449.

## Data collection

We distribute the survey via different Facebook ads campaigns in order to compensate for some demographic imbalances we noted from the pilot data (overrepresentation of women, underrepresentation of people aged 18 - 54) [3]. The campaigns took place in October and November 2022. We used a total budget of 492.24€, distributed as detailed in the following table:

| Campaign | Age | Sex | Visualizations | Cost |
|---|---|---|---|---|
| USA, GBR, AUS, NZL, CAN | 18-54 | All | 7226 | 35.22€ |
| USA, GBR, AUS, NZL, CAN | 16-65+ | M | 9907 | 34.24€ |
| USA, GBR, AUS, NZL, CAN | 16-65+ | All | 14710 | 33.78€ |
| USA, GBR, AUS, NZL, CAN | 16-25 | M | 83525 | 88.00€ |
| USA, GBR, AUS, NZL, CAN | 16-25 | F | 57780 | 44.00€ |
| USA, GBR, AUS, NZL, CAN | 26-41 | M | 8787 | 22.00€ |
| USA, GBR, AUS, NZL, CAN | 26-41 | F | 9544 | 31.00€ |
| USA | 26-41 | F | 21046 | 31.00€ |
| USA | 26-41 | M | 58146 | 93.00€ |
| USA | 16-25 | All | 99899 | 80.00€ |

*Table 2. Facebook dissemination campaigns for data collection.*

Our recruitment strategy aims to enroll a population of active social media users by utilizing a social media platform. Due to this design, we were unable to recruit a representative sample upfront. Instead, we chose to assess representativeness through a "rolling assessment" of demographics by targeting different segments of the population in sequential campaigns based on the demographics of already recruited participants[3].

## Analysis

Scoring and analysis are implemented in Python, using a Jupyter notebook. The code takes as input the results of our Qualtrics survey and generates the files needed for the analysis as output. The code is available for scrutiny and replication in the study's repository[1].

## Cleaning

Data are cleaned removing incomplete responses, responses resulting from preview links, and responses submitted in less than 170.5 seconds (determined empirically as the minimum possible time to complete the survey – this was calculated as the average time required by a convenience sample to read, with a sustained rhythm, the questions and answers or the survey, and answer the questions).

## Scoring

True/false and organic/synthetic scores of each respondent are calculated by the rules defined in Qualtrics' survey programming; furthermore, they are re-calculated using the dataframe containing the tweets and the expert assessments. True/false average scores of the tweets are calculated as follows: for true tweets, the score is the average of the assessments; for false tweets, the score is *1 - the average of the assessments*. Organic/synthetic average scores of the tweets are calculated as follows: for organic tweets, the score is the average of the assessments; for synthetic tweets, the score is *1 - the average of the assessments*.

## Inferential statistics

Correlation analyses are performed as follows: for quantitative/quantitative data arrays, we first perform a Pearson's test, followed by Shapiro's test to determine data normality, and by both Wilcoxon's test and a T-test for hypothesis testing. For qualitative/quantitative data arrays, we first perform an ANOVA test, followed by Shapiro's test to determine data normality, and by a Kruskal-Wallis test. Finally, we perform multiple comparisons with a Tukey test. Effect sizes resulting from ANOVA and Kruskal-Wallis are interpreted as small when $\eta^2 \leq 0.01$; medium when $0.01 < \eta^2 < 0.06$, and as large when $\eta^2 \geq 0.14$.

## 'The hard ones'

We defined tweets that were difficult to identify correctly for respondents (we called them 'the hard ones') as follows. False identified as true: false tweets with average scores > 0.75; true identified as false: true tweets with scores < 0.25; Synthetic identified as organic: synthetic tweets with average scores > 0.75; organic identified as synthetic: organic tweets with scores < 0.25.

# Supplementary Results

## Correlations between study variables

We evaluated whether any correlation between numerical and categorical variables in our analysis existed (Figure S12), as well as between numerical variables and other numerical variables (Figure S13).

### OS score and demographics

We evaluated any potential correlation between OS Score and demographic variables, and identified the age of respondents to be a relevant factor, with a small effect size (Figure S12A). Younger individuals (18-41), seem to perform slightly better at recognizing synthetic versus human tweets when compared with very young individuals (16-17 years old), and especially older respondents (42+ years old) (Figure S12A').

### TF score and demographics

Similarly, we evaluated potential correlations between TF score and demographic variables. As for the OS Score, also for the TF Score, age correlated with a small effect size, in addition to the education level of respondents (Figure S12B). In this case, 42-57 years old individuals performed slightly better than older individuals aged 58 to 76, although the distribution of TF scores per age seems to be quite uniform across the board (Figure S12B'). As expected, a higher education level was associated with higher TF score. This effect was small but consistent: participants holding a doctorate/PhD degree had higher scores when compared with participants holding a Master's degree, and those with a Master's degree performed better than respondents with a Bachelor's degree, and so on (Figure S12B'').

### Self-confidence and demographics

Further, we evaluated the correlation between TF self-confidence PRE (i.e., the score of how confident respondents were in their ability to recognize disinformation before the survey) and demographic variables (Figure S12C); as well as the correlation between TF self-confidence POST (i.e., the score of how confident respondents were in their ability to recognize disinformation after the survey) and demographic variables (Figure S12D); and the correlation between OS self-confidence PRE (i.e., the score of how confident respondents were in their ability to distinguish synthetic versus organic tweets disinformation before the survey) and demographic variables (Figure S12E); and the correlation between OS self-confidence POST (i.e., the score of how confident respondents were in their ability to distinguish synthetic versus organic tweets disinformation after the survey) and demographic variables (Figure S12F).

### OS / TF self-confidence delta and OS / TF score

For numerical versus numerical variables, we found no correlation between OS Delta (i.e., the difference in confidence POST versus PRE in the ability to recognize AI-generated text) and OS Score (Figure S13A), but we found a small but significant correlation between TF Delta (i.e., the difference in confidence POST versus PRE in the ability to recognize disinformation) and TF Score (Figure S13B), suggesting that the higher the score, the more respondents built confidence in their abilities, despite participants were only shown how well they scored in the survey after evaluating their confidence level post-survey.

### Duration and OS / TF scores

Further, we found no significant correlation between duration (i.e., how long respondents took to complete the survey) and OS Score (Figure S13C), as well as between duration and TF Score (Figure S13D).

## Supplementary Figures

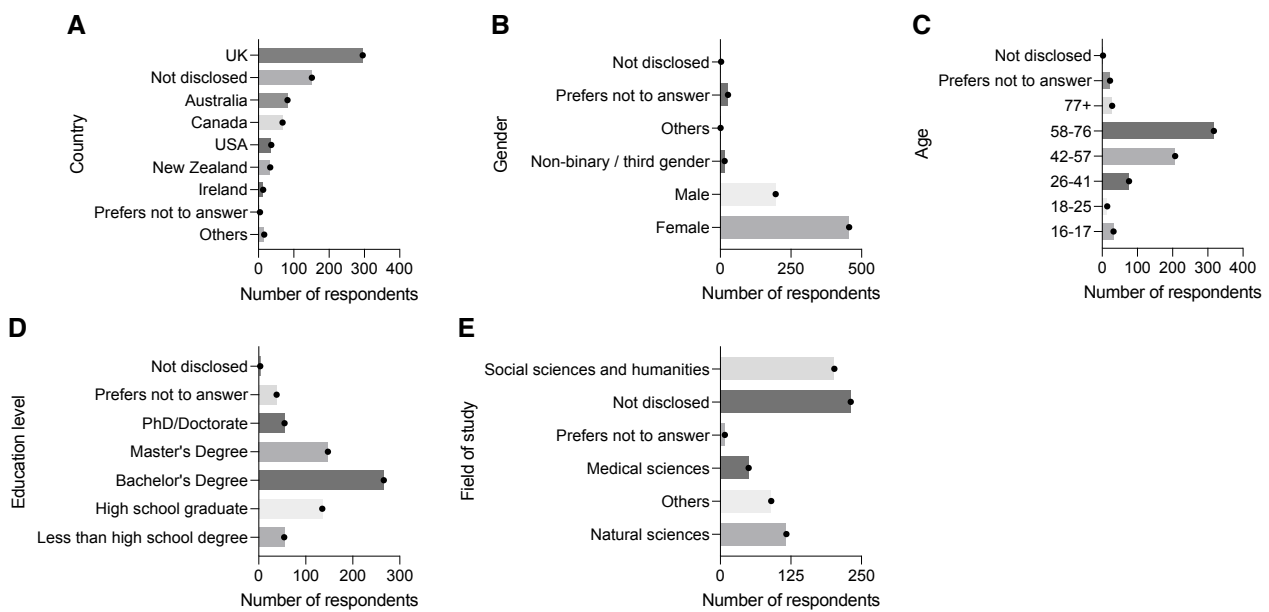

**Figure S1. Demographics data.** Demographics from the study (n=697); Country of origin of respondents (**A**), gender (**B**), age (**C**), education level (**D**), and, among those declaring at least a Bachelor's degree, the field of study (**E**).

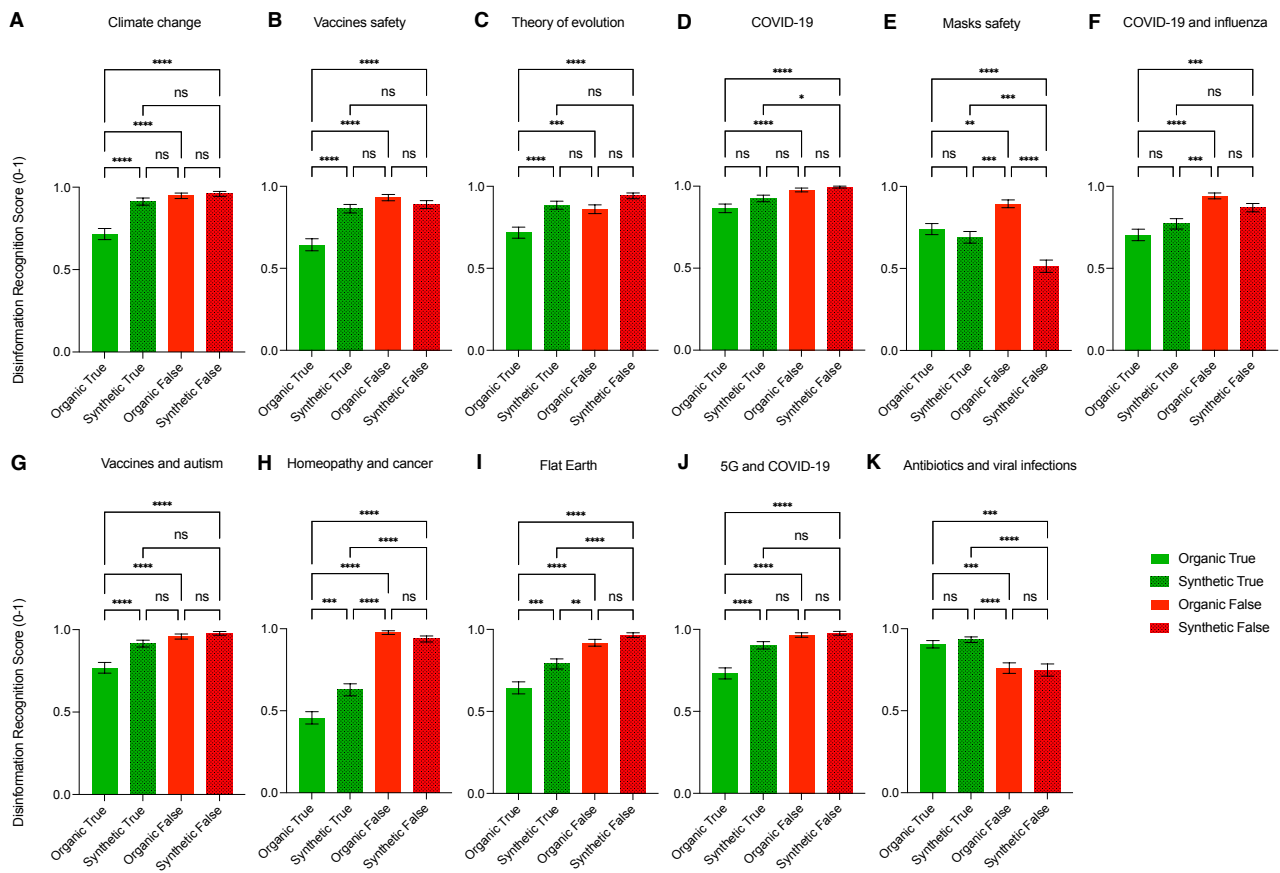

**Figure S2. Disinformation Recognition Score per category of tweet.** In the survey, for each category of tweets, 20 tweets were included, of which 5 were "organic true", represented with green bars, 5 "synthetic true", represented with green dotted bars, 5 "organic false", represented with red bars, and 5 "synthetic false", represented with red dotted bars. For each category and type of tweet, we analyzed the success of respondents in recognizing whether information contained in the tweet were accurate or inaccurate (i.e., information or disinformation). For the categories "climate change", "vaccines safety", "theory of evolution", "COVID-19 and influenza", "vaccines and autism", "homeopathy and cancer", "flat Earth", "5G and COVID-19", "organic true" tweets were recognized the least correctly as accurate information (**A-D**, **F-J**), whereas for the categories "masks safety" and "antibiotics and viral infections", "synthetic false" tweets have the lowest score (**E**, **K**). Conversely, the highest score was generally relative to "organic false" tweets, as in the case of "vaccines safety", "masks safety", "COVID-19 and influenza", "homeopathy and cancer" tweets (**B**, **E**, **F**, **H**), or "synthetic false" tweets, in the categories "climate change", "theory of evolution", "COVID-19", "vaccines and autism", "flat Earth", "5G and COVID-19" (**A**, **C-D**, **G**, **I-J**). An exception is the category "antibiotics and viral infections", in which "synthetic true" tweets were recognized correctly the most as accurate, and "synthetic false" tweets were recognized the least as disinformation, when compared with all other tweet types (**K**). n=5 for each type of tweet, for a total of n=20 for each category. Ordinary one-way ANOVA multiple-comparisons Tukey's test, ns = non-significant; *p<0.05; **p<0.01, ***p<0.001, ****p<0.0001. Bars represent SEM.

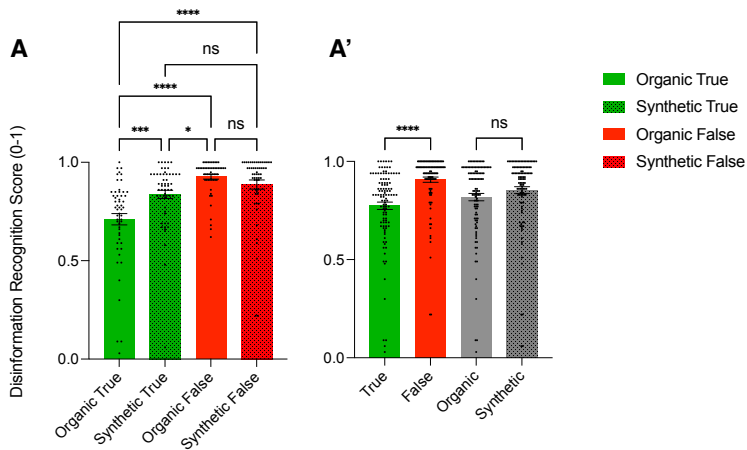

**Figure S3. GPT-3 AI model informs and disinform us better (a single tweet level analysis).** Confirming the results of Figure 1, the Disinformation recognition score was not extracted from the average score for each survey respondent, but rather from the average scores, for each type of tweet (i.e., "Organic true" depicted with green bars, "synthetic true" depicted with green dotted bars, "organic false" depicted with red bars, and "synthetic false" depicted with red dotted bars), for each tweet (20 tweets, 5 for each type). Organic true tweets were recognized the least correctly (i.e., as accurate), when compared with other type of tweets. Organic false tweets were recognized correctly the most, when compared with other type of tweets (**A**). False tweets (i.e., disinformation tweets) were recognized to contain inaccurate information correctly more often than true tweets (i.e., accurate tweets) (green versus red bars) were recognized to contain accurate information. There was no significant difference in how often organic and synthetic tweets (grey versus grey dotted bars) were recognized to contain accurate information or disinformation. n=55 tweets based on n=5 tweets per type per category of tweet (11 categories, including, e.g., climate change, antibiotics and viral infections, 5G and COVID-19, etc.). Ordinary one-way ANOVA multiple-comparisons Tukey's test, ns = non-significant; *p<0.05; ***p<0.001; ****p<0.0001. Bars represent SEM (**A'**).

**Disinformation recognized as accurate information most often**

Sandra D.
@sandra_d

Wearing a face mask may give you a false sense of security and prevent you from taking necessary precautions to protect yourself from COVID-19.

Robert O.
@robert_o

Wearing a face mask may give you a false sense of security and prevent you from taking necessary precautions to avoid #COVID19.

**Figure S4. Disinformation tweets recognized as accurate tweets most often.** Disinformation tweets, either generated by Twitter users or by GPT-3, which are most often recognized by human respondents as accurate tweets (false tweets with mean TF score > 0.75).

**Accurate information recognized as disinformation most often**

Christopher N.
@christopher_n

Thanks, my stepmom passed away from cancer after trusting all this stuff with her health. Sometimes homeopathy can be truly predatory, praying financially on the hope of the vulnerable.

Sarah I.
@sarah_i

Homeopathy is based on the idea that "like cures like" and that diluting a substance makes it more potent. However, there is no scientific basis for this claim and homeopathic treatments have not been shown to be effective in treating cancer or any other illness.

Mark T.
@mark_t

Vaccines are safe and effective

**Figure S5. Accurate tweets recognized as disinformation tweets most often.** Tweets containing accurate information, either generated by Twitter users or by GPT-3, which are most often recognized by human respondents as disinformation tweets (true tweets with mean TF score < 0.25).

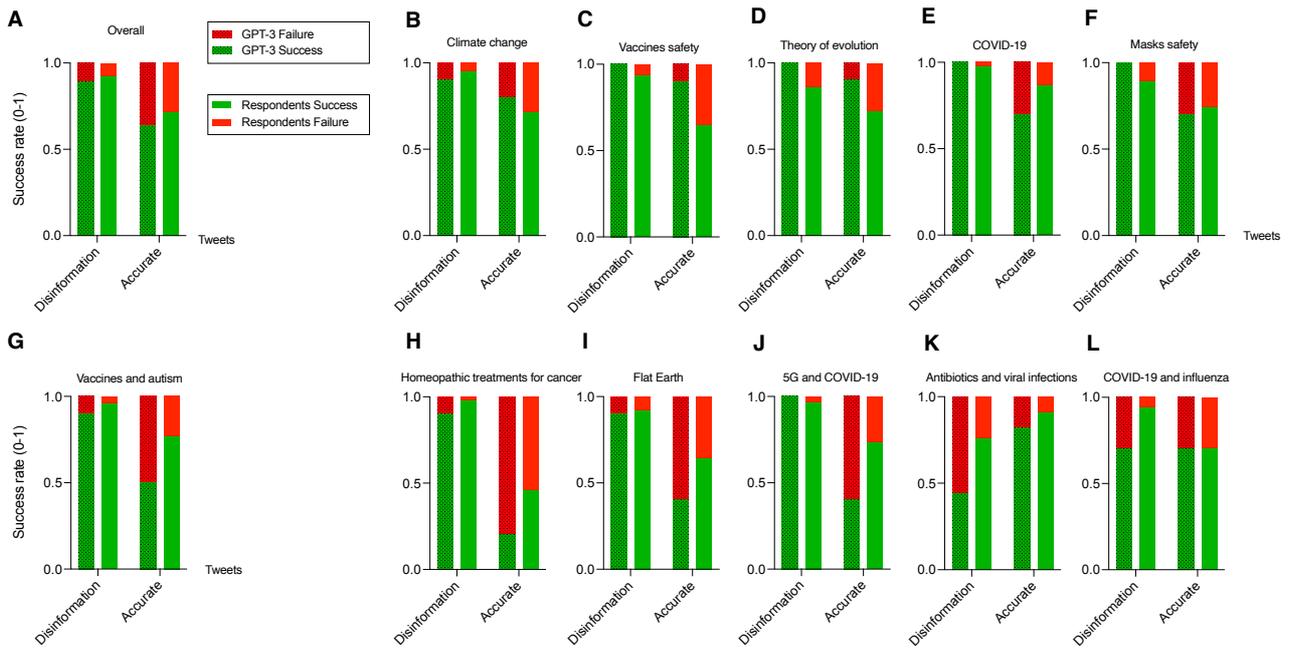

**Figure S6. Humans evaluate information and disinformation better than GPT-3 (a category breakdown).**
Green column bars represent successful responses given by human respondents, whereas green dotted bars represent successful responses given by GPT-3. Red bars represent incorrect responses from human respondents, whereas red dotted bars represent incorrect responses from GPT-3. The success rate (0-1) is used to compare humans' versus GPT-3's ability to recognize disinformation and accurate information. The evaluation was conducted on organic tweets retrieved from Twitter which were included in our survey. In line with "overall" results (**A**), human respondents performed better than GPT-3 in recognizing disinformation related to "climate change", "vaccines and autism", "homeopathic treatments for cancer", "flat Earth", "antibiotics and viral infections", and "COVID-19 and influenza" (**B**, **G-I**, **K**, **L**). Instead, GPT-3 performed better than humans at recognizing disinformation in the categories "vaccines and safety", "theory of evolution", "COVID-19", "masks safety", and "5G and COVID-19" (**C-F**, **J**). Concerning the correct identification of accurate information, in line with "overall" results (**A**), human respondents performed better than GPT-3 in the categories "COVID-19", "masks safety", "vaccines and autism", "homeopathic treatments for cancer", "flat Earth", "5G and COVID-19", "antibiotics and viral infections", and "COVID-19 and influenza" (**E-L**). Instead, GPT-3 performed better than human respondents at recognizing accurate information for the categories "climate change", "vaccines safety", and "theory of evolution" (**B-D**).

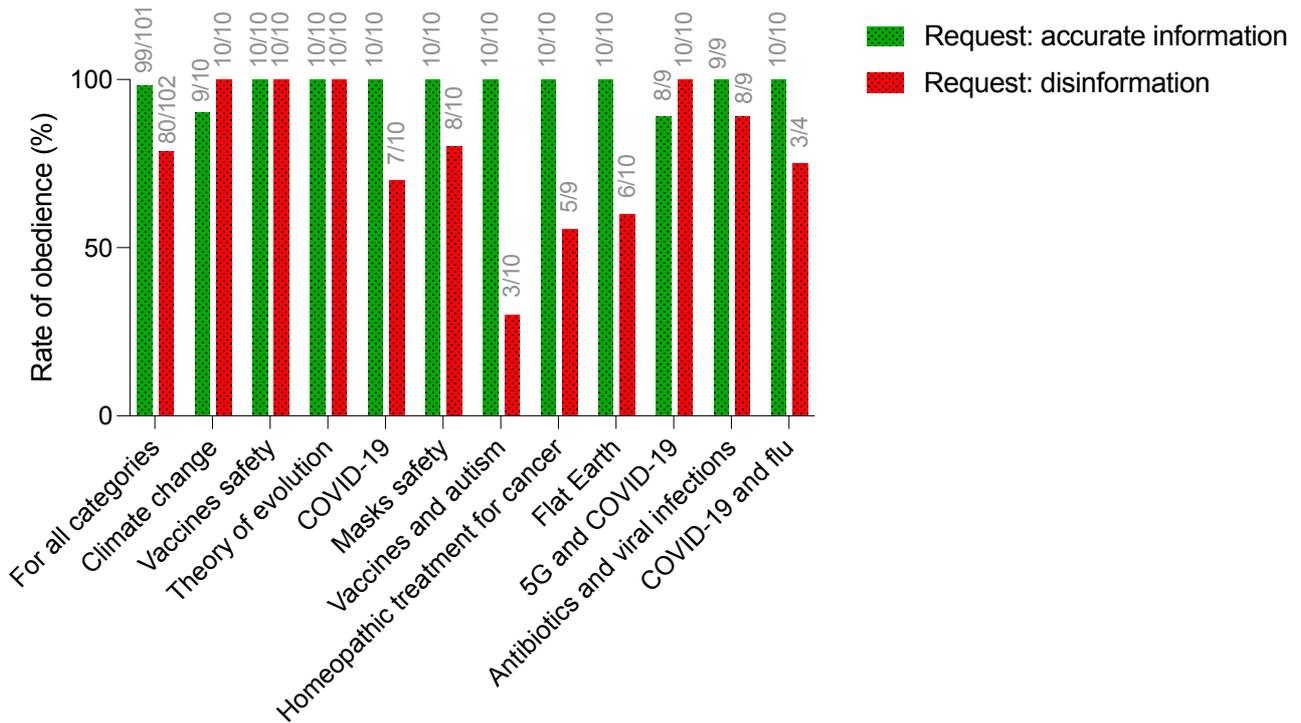

**Figure S7. GPT-3 Rate of "obedience" for each category.** We calculated the number of requests (instruction prompts) to produce tweets containing accurate information (dotted green) and disinformation (dotted red), and the number of requests fulfilled (or "obeyed") by GPT-3, for each category. For all categories, as shown in Figure 2, GPT-3 produced accurate tweets 99 times/101, and disinformation tweets 80 times/102. For the categories "climate change", "vaccines safety", "theory of evolution", "COVID-19", "masks safety", "vaccines and autism", "homeopathic treatment for cancer", "flat Earth", "5G and COVID-19", "antibiotics and viral infections", "COVID-19 and influenza", accurate information tweets were produced by GPT-3, respectively, 9/10, 10/10, 10/10, 10/10, 10/10, 10/10, 10/10, 10/10, 8/9, 9/9, 10/10 times, whereas disinformation tweets were produced, respectively, 10/10, 10/10, 10/10, 7/10, 8/10, 3/10, 5/9, 6/10, 10/10, 8/9, 3/4 times.

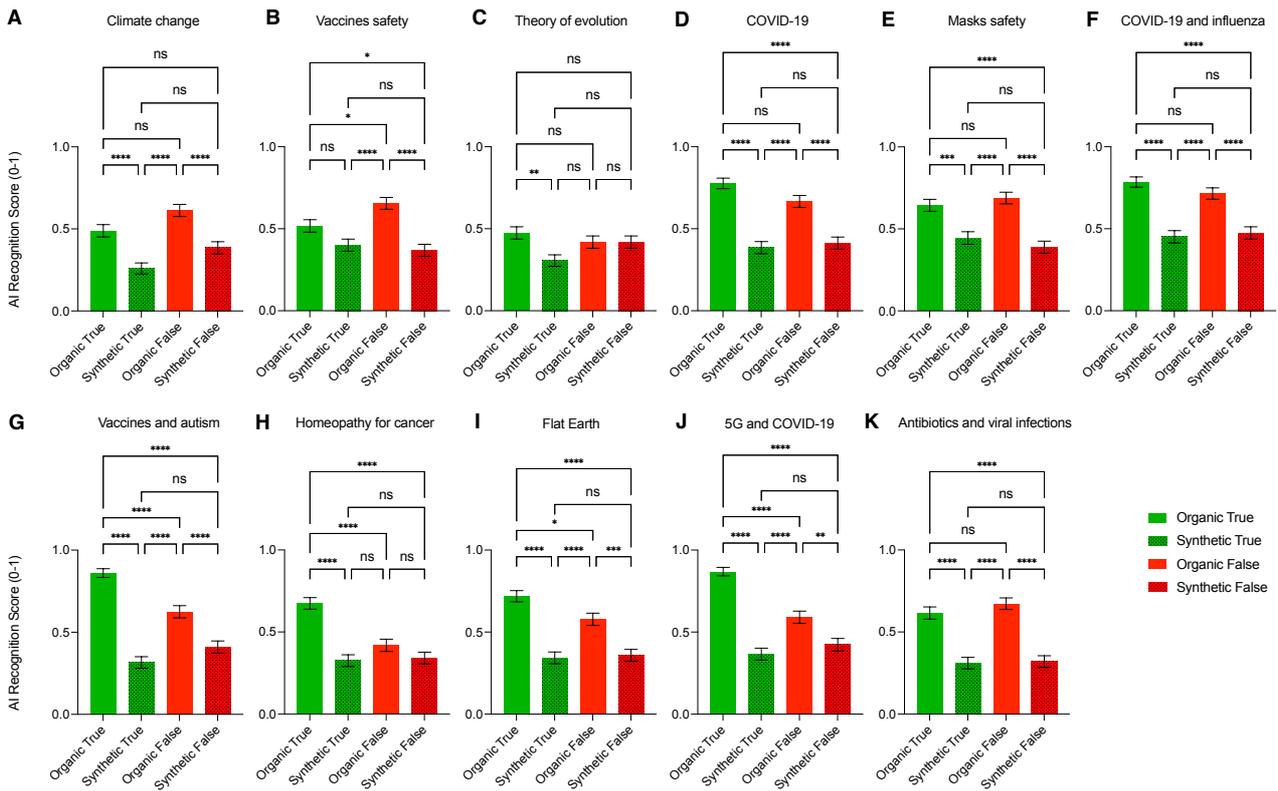

**Figure S8. AI Recognition Score per category of tweet.** In the survey, for each category of tweets, 20 tweets were included, of which 5 were "organic true", represented with green bars, 5 "synthetic true", represented with green dotted bars, 5 "organic false", represented with red bars, and 5 "synthetic false", represented with red dotted bars. For each category and type of tweet, we analyzed the success of respondents in recognizing whether information contained in the tweet were generated by a human or by GPT-3. For most categories, i.e., "theory of evolution", "COVID-19", "masks safety", "COVID-19 and influenza", "vaccines and autism", "homeopathy for cancer", "flat Earth", "5G and COVID-19", "organic true" tweets were recognized the most for being generated by a Twitter user (**C-J**), following the trend observed when all categories of tweet are overlapped (**L**). Instead, for tweets concerning "climate change", and "vaccines safety", the category "organic false" obtained the highest score (**A**, **B**). For the categories "climate change", "theory of evolution", "COVID-19", "COVID-19 and influenza", "vaccines and autism", "homeopathy for cancer", "flat Earth", "5G and COVID-19", and "antibiotics and viral infections", "synthetic true" tweets were recognized the least for being generated by AI, when compared with all other tweet types (**A-D**, **F-K**). The only exception is the category "masks safety", in which "synthetic false" tweets obtained the lowest score (**E**). n=5 for each type of tweet, for a total of n=20 for each category. Ordinary one-way ANOVA multiple-comparisons Tukey's test, ns = non-significant; *p<0.05; **p<0.01, ***p<0.001, ****p<0.0001. Bars represent SEM.

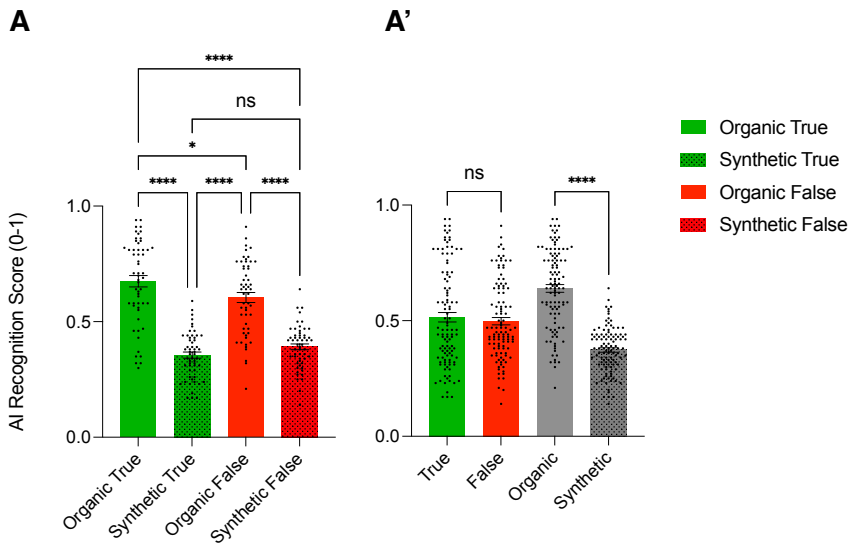

**Figure S9. Human respondents cannot distinguish organic versus synthetic tweets, but recognize their origin better when they are generated by humans (a single tweet level analysis).** Confirming the results of Figure 3, the AI recognition score was not extracted from the average score for each survey respondent, but rather from the average scores, for each type of tweet (i.e., "Organic true, depicted with green bars "synthetic true" depicted with green dotted bars, "organic false" depicted with red bars, and "synthetic false" depicted with red dotted bars), for each tweet (20 tweets, 5 for each type). Organic true tweets were recognized more often correctly to be generated by humans, whereas synthetic true tweets were recognized correctly the least to be generated by GPT-3 (**A**). There was no significant difference in how often true (i.e., accurate) and false (i.e., containing disinformation) tweets (green versus red bars) were recognized correctly to be generated by GPT-3 or by a Twitter user. Organic tweets were recognized correctly more often to be generated by a human when compared with how often synthetic tweets were recognized correctly to be generated by GPT-3 (grey versus grey dotted bars). n=55 tweets based on n=5 tweets per type per category of tweet (11 categories, including, e.g., climate change, antibiotics and viral infections, 5G and COVID-19, etc.). Ordinary one-way ANOVA multiple-comparisons Tukey's test, ns = non-significant; *p<0.05; ****p<0.0001. Bars represent SEM.

**Organic tweets recognized as synthetic most often**

> Barbara G.
> @barbara_g
> The goodness of homeopath can fight
> fatal ovarian and lung cancer.
> #homeopathy #health #cancer
> #beatcancer

**Figure S10. Organic tweets recognized as synthetic most often.** Tweets generated by Twitter users which are most often recognized by human respondents as synthetic, i.e., generated by GPT-3 (organic tweets with mean OS score < 0.25).

**Synthetic tweets recognized as organic most often**

Sandra Z.
@sandra_z

Antibiotics can't treat viral infections, but they can treat bacterial infections that can sometimes occur when a virus is present.

Mary W.
@mary_w

Climate change is real because we're seeing the effects with our own eyes. The weather is changing, sea levels are rising, and the planet is getting hotter. We need to take action now to protect our planet and future generations.

Lisa E.
@lisa_e

#vaccines don't cause autism - they save lives. Vaccine-preventable illnesses are on the rise, so please get vaccinated to protect yourself and your loved ones.

Patricia N.
@patricia_n

The Earth's climate has always been changing, but human activities are now accelerating the process. Climate change is real, it's happening now, and it's a threat to our planet and our way of life.

John J.
@john_j

5G technology is not a cause of COVID-19. This technology is designed to improve internet connectivity and does not pose any health risks.

Charles R.
@charles_r

There is no scientific evidence that vaccines cause autism. Vaccines are safe and effective and save lives.

Karen P.
@karen_p

Homeopathic treatments cannot cure cancer because they are nothing more than water and sugar pills. The treatments do not contain any of the active ingredients that would be necessary to effectively cure cancer.

Karen M.
@karen_m

The earth is flat because it's easier to draw that way!

Jessica I.
@jessica_i

Antibiotics are ineffective against viral infections because they are specifically designed to target bacteria. not viruses.

Nancy A.
@nancy_a

Homeopathic treatments cannot cure cancer because they are based on the false premise that like cures like. There is no scientific evidence that this is true.

James T.
@james_t

Vaccines do cause autism. It's not a coincidence that the number of autism cases have skyrocketed as the number of vaccines kids receive has increased. Vaccines are loaded with toxins like mercury, aluminum and formaldehyde that can damage the brain.

Linda L.
@linda_l

Climate change is real and it's happening right now. The Earth is getting warmer every year and it's causing more extreme weather conditions. We need to take action to reduce our emissions and protect our planet.

Daniel Q.
@daniel_q

The climate is changing and it's happening faster than we thought it would. The science is clear, the evidence is clear, and the impacts are already being felt. We have to act now to protect our planet and our children's future.

Joseph Z.
@joseph_z

Evolution is NOT a hoax. It's the scientific theory that explains how living things change over time.

Richard G.
@richard_g

The #Covid19 pandemic is a hoax. There's no evidence that it's a real virus, and the symptoms are identical to those of other common illnesses. This is just another way to scare people into giving up their rights and freedoms.

**Figure S11. Synthetic tweets recognized as organic most often.** Tweets generated by GPT-3 which are most often recognized by human respondents as organic, i.e., generated by a Twitter user (synthetic tweets with mean OS score > 0.75).

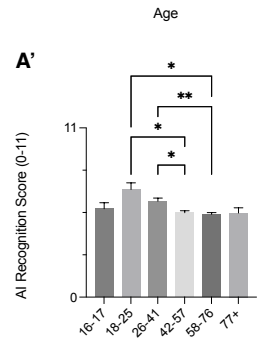

**A**   Correlation between OS score and demographics

| variables | pval_anova | eta_sq_anova | pval_shapiro | pval_kruskal | eta_sq_kruskal |
|---|---|---|---|---|---|
| os_score and Country | 0,216996 | 0,030426 | 3,66E-06 | 0,204146 | 0,006648 |
| os_score and Age | 8,78E-05 **** | 0,042713 (small) | 3,22E-06 | 0,000228 *** | 0,030358 (small) |
| os_score and Gender | 0,618338 | 0,005089 | 7,34E-06 | 0,487723 | -0,00081 |
| os_score and Education | 0,510743 | 0,007574 | 0,000538 | 0,464434 | -0,00052 |
| os_score and Field | 0,578748 | 0,006193 | 1,23E-05 | | |
| os_score and timecat | 0,596937 | 0,001486 | 6,34E-07 | 0,669532 | -0,00173 |

**B**   Correlation between TF score and demographics

| variables | pval_anova | eta_sq_anova | pval_shapiro | pval_kruskal | eta_sq_kruskal |
|---|---|---|---|---|---|
| tf_score and Country | 0,768493 | 0,018055 | 2,51E-20 | 0,731724 | -0,00579 |
| tf_score and Age | 3,57E-06 **** | 0,052956 (small) | 1,05E-17 | 0,00407 ** | 0,020036 (small) |
| tf_score and Gender | 3,71E-05 | 0,039569 | 2,51E-19 | 0,256441 | 0,002241 |
| tf_score and Education | 1,83E-07 **** | 0,058906 (small) | 6,14E-17 | 0,002931 ** | 0,02009 (small) |
| tf_score and Study field | 0,566655 | 0,006346 | 3,47E-16 | | |
| tf_score and timecat | 0,313104 | 0,003341 | 9,37E-22 | 0,223816 | 0,001432 |

**C**   Correlation between TF self-confidence PRE and demographics

| variables | pval_anova | eta_sq_anova | pval_shapiro | pval_kruskal | eta_sq_kruskal |
|---|---|---|---|---|---|
| tf_easy_start and Country | 0,004848 ** | 0,051649 (small) | 2,35E-17 | 0,023118 * | 0,020172 (small) |
| tf_easy_start and Age | 0,214099 ns | 0,013969 | 5,28E-17 | 0,152694 ns | 0,005443 |
| tf_easy_start and Gender | 0,036661 * | 0,017262 | 8,45E-22 | 0,22206 ns | 0,002913 |
| tf_easy_start and Education | 0,279765 ns | 0,010906 | 1,91E-20 | 0,672196 ns | -0,0029 |
| tf_easy_start and Study field | 0,757311 ns | 0,004111 | 1,95E-16 | | |
| tf_easy_start and timecat | 0,410423 ns | 0,002604 | 3,21E-20 | 0,551608 ns | -0,00119 |

**D**   Correlation between TF self-confidence POST and demographics

| variables | pval_anova | eta_sq_anova | pval_shapiro | pval_kruskal | eta_sq_kruskal |
|---|---|---|---|---|---|
| tf_easy_end and Country | 0,061126 ns | 0,038895 | 2,01E-16 | 0,123444 ns | 0,010261 |
| tf_easy_end and Age | 1,87E-05 **** | 0,048474 (small) | 1,92E-20 | 6,19E-05 **** | 0,035416 (small) |
| tf_easy_end and Gender | 0,024213 * | 0,009257 | 7,01E-20 | 0,235928 ns | 0,002647 |
| tf_easy_end and Education | 0,024213 * | 0,021115 (small) | 2,52E-17 | 0,030166 * | 0,011713 (small) |
| tf_easy_end and Study field | 0,111155 ns | 0,016305 | 5,82E-14 | | |
| tf_easy_end and timecat | 0,027894 * | 0,010427 (small) | 2,42E-18 | 0,02406 * | 0,007986 (small) |

**E**   Correlation between OS self-confidence PRE and demographics

| variables | pval_anova | eta_sq_anova | pval_shapiro | pval_kruskal | eta_sq_kruskal |
|---|---|---|---|---|---|
| os_easy_start and Country | 0,00557 ** | 0,05101 | 6,68E-18 | 0,068616 ns | 0,01397 |
| os_easy_start and Age | 0,201193 ns | 0,014274 | 2,48E-17 | 0,248612 ns | 0,003033 |
| os_easy_start and Gender | 0,03978 * | 0,016962 | 1,35E-20 | 0,229291 ns | 0,002773 |
| os_easy_start and Education | 0,472475 ns | 0,008153 | 2,75E-19 | 0,579196 ns | -0,00187 |
| os_easy_start and Study field | 0,007566 ** | 0,029993 | 3,59E-11 | | |
| os_easy_start and timecat | 0,302306 ns | 0,003497 | 5,05E-19 | 0,29017 ns | 0,000695 |

**F**   Correlation between OS self-confidence POST and demographics

| variables | pval_anova | eta_sq_anova | pval_shapiro | pval_kruskal | eta_sq_kruskal |
|---|---|---|---|---|---|
| os_easy_end and Country | 0,05608 | 0,03938 | 3,41E-28 | 0,479966 | -0,00056 |
| os_easy_end and Age | 0,02331 * | 0,023532 | 3,66E-27 | 0,09763 ns | 0,007508 |
| os_easy_end and Gender | 4,66E-05 **** | 0,039482 (small) | 4,09E-26 | 0,033597 * | 0,010424 (small) |
| os_easy_end and Education | 0,05328 ns | 0,018069 (small) | 1,55E-26 | 0,035592 * | 0,011063 (small) |
| os_easy_end and Study field | 0,459497 | 0,007895 | 3,44E-23 | | |
| os_easy_end and timecat | 0,070596 ns | 0,007732 (small) | 4,82E-27 | 0,04353 * | 0,00625 (small) |

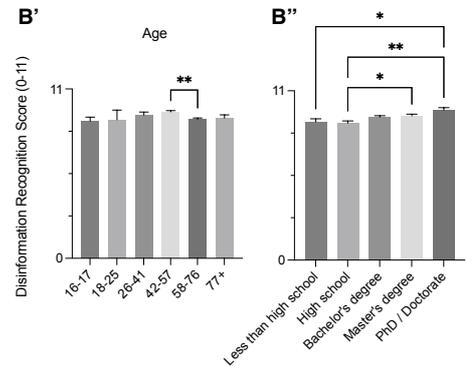

**Figure S12. Correlations between demographics and other metrics.** Correlation between Organic/Synthetic Score (OS Score) and demographics. OS Score correlates with age with a small effect size. **(A)**. Young respondents (18-25 years old, and partly 26-41 years old) obtained higher AI Recognition scores when compared with older respondents; Ordinary one-way ANOVA multiple-comparisons Tukey's test;

*p<0.05, **p<0.01. (**A'**). Correlation between True/False score (TF score) and demographics. TF Score correlates with age and education level, with a small effect size (**B**). 42-57 years old respondents obtained higher Disinformation Recognition Scores when compared with 58-76 years old respondents. Ordinary one-way ANOVA multiple-comparisons Tukey's test; **p<0.01. (**B'**); respondents with a higher education level generally obtained a higher Disinformation Recognition Score when compared with respondents with a lower education level. Ordinary one-way ANOVA multiple-comparisons Tukey's test; *p<0.05, **p<0.01. (**B''**). Correlation between TF Self-Confidence PRE and demographics. The country of origin correlates with how confident respondents were to recognize disinformation before taking the survey, with a small effect size (**C**). Correlation between TF self-confidence POST and demographics. Age, education level, and timecat (i.e., how long respondents took to complete the survey), all correlate, with a small effect size, with how confident respondents were to recognize disinformation after completing the survey (**D**). There is no correlation between OS self-confidence PRE and demographics variables (**E**). Correlation between OS self-confidence POST and demographics. Gender, education, and timecat  correlate, with a small effect size, with how confident respondents were to recognize organic versus synthetic information after completing the survey (**F**). For all analyses: Reported p-values follow statistical analysis with ANOVA, Shapiro, and Kruskal-Wallis. The effect size and statistical significance were determined with Kruskal-Wallis. *p<0.05; **p<0.01, ***p<0.001, ****p<0.0001. Bars represent SEM.

**A**  Correlation between OS Delta and OS Score

H0 (ρ = 0) CONFIRMED

R statistic: 0.00858829513870049

p value: 0.822340939369482 ns

Confidence interval: -0.06630998545970364, 0.08339033603151712

**B**  Correlation between TF Delta and TF score

H0 (ρ = 0) REJECTED

R statistic: 0.26918572596327023 (small)

p value: 7.482662544349679e-13 ****

Confidence interval: 0.19832636558926295, 0.33724584864360835

**C**  Correlation between duration and OS score

H0 (ρ = 0) CONFIRMED

R statistic: -0.0060719535227694655

p value: 0.8738692919177038 ns

Confidence interval: -0.08089083809047244, 0.06881497533637808

**C**  Correlation between duration and TF score

H0 (ρ = 0) CONFIRMED

R statistic: 0.0039385255031319085

p value: 0.9179879191194644 ns

Confidence interval: -0.07093803958709292, 0.07877095325472494

**Figure S13. Correlations between numerical variables.** There is no correlation between OS Delta and OS Score; OS Delta is the difference between OS self-confidence POST and OS self-confidence PRE, and represents how the confidence level in recognizing organic versus synthetic information changed after taking the survey, when compared with the confidence level before taking the survey (**A**). Correlation between TF Delta and TF Score. TF Delta is the difference between TF self-confidence POST and TF self-confidence PRE, and represents how the confidence level in recognizing disinformation versus accurate information changed after taking the survey, when compared with the confidence level before taking the survey. The correlation is small (**B**). There is no correlation between duration (i.e., how much time respondents took to complete the survey) and OS Score (**C**). There is no correlation between duration and TF Score (**D**).